\documentclass[runningheads]{llncs}
\usepackage{url}

\DeclareMathAlphabet{\mathrm}{OT1}{cmr}{m}{n}
\DeclareMathAlphabet{\mathrmbf}{OT1}{cmr}{bx}{n}
\DeclareMathAlphabet{\mathrmit}{OT1}{cmr}{m}{it}
\DeclareMathAlphabet{\mathrmbfit}{OT1}{cmr}{bx}{it}
\DeclareMathAlphabet{\mathsf}    {OT1}{cmss}{m}{n}
\DeclareMathAlphabet{\mathsfbf}  {OT1}{cmss}{bx}{n}
\DeclareMathAlphabet{\mathsfit}  {OT1}{cmss}{m}{sl}
\DeclareMathAlphabet{\mathttbf}{OT1}{cmtt}{bx}{n}
\newcommand{\keywords}[1]{\par\addvspace\baselineskip\noindent\enspace\ignorespaces{\bfseries Keywords:\,}#1}
\newenvironment{principle}[1]{\begin{flushleft}\normalsize\em{{\begin{bfseries}\begin{upshape} #1 Principle:\end{upshape}\end{bfseries}}\/}}
{\end{flushleft}}
\begin{document}

\title{Information Flow in Logical Environments}
\author{Robert E. Kent}
\institute{Ontologos}
\maketitle

\begin{abstract}
This paper describes information flow within logical environments.
The theory of information flow, the logic of distributed systems,
was first defined by Barwise and Seligman \cite{barwise:seligman:97}.
Logical environments are a semantic-oriented version of institutions.
The theory of institutions,
which was initiated by Goguen and Burstall \cite{goguen:burstall:92}, 
is abstract model theory.
Information flow is the flow of information in channels over distributed systems.
The semantic integration of distributed systems,
be they ontologies, databases or other information resources, 
can be defined in terms of the channel theory of information flow.
As originally defined,
the theory of information flow uses only a specific logical environment 
in order to discuss information flow.
This paper shows how information flow can be defined in an arbitrary logical environment.
\keywords{information flow, logical environment, distributed systems, channel, core, information integration, structure, theory and logic}
\end{abstract}


\section{Introduction}

We describe information flow in logical environments.
The theory of Information Flow ({\ttfamily IF}) 
is \cite{barwise:seligman:97} the logic of distributed systems.
The theory of Institutions ({\ttfamily INS})
is \cite{goguen:burstall:92} abstract model theory.
Both can be regarded 
as ways to describe, classify, model, extract, and apply patterns of knowledge
in the design and maintenance of 
ontologies \cite{kent:iswc03}, databases \cite{goguen:05} and other knowledge resources.
The theory {\ttfamily IF} and the theory {\ttfamily INS} are related in two ways.
On the one hand,
{\ttfamily IF} is a special case of first order logic ({\ttfamily FOL}) 
(and {\ttfamily FOL} is just one institution):
the types of {\ttfamily IF} can be regarded as unary predicates and (then)
the instances of {\ttfamily IF} are to be regarded as individuals
that may satisfy these unary predicates.
So {\ttfamily IF} is both a special case and a specific instance of {\ttfamily INS}.
This point of view,
advocated in a recent paper by Goguen \cite{goguen:06},
has been used for several years in the Information Flow Framework ({\ttfamily IFF}),
as evidenced by the various documents and ontologies 
listed at the {\ttfamily IFF} website \cite{url}.
On the other hand,
the {\ttfamily IF} classification relation (between instances and types)
can be regarded as an abstraction of 
the {\ttfamily INS} satisfaction relation (between structures and sentences).
As such,
anything in the theory of {\ttfamily IF} that is treated abstractly
can be used to extend the theory of {\ttfamily INS}. 
So we can view {\ttfamily IF} as an abstraction and extension of {\ttfamily INS}.
This paper argues for this point of view.
We first discuss a more semantic form of institutions called logical environments.
Then we demonstrate how the basics of information flow can be define in logical environments.
We follow the example set by the theory of {\ttfamily IF} \cite{barwise:seligman:97} 
in the use of principles to codify and structure the discussion.

\section{Logical Environments}

\subsection{Structures}

\begin{principle}{World}
The world exists a priori and is describable.
\end{principle}

\noindent
We represent the world as a category (mathematical context)
$\mathrmbf{Struc}$ of semantical structures and structure morphisms.
We assume the category of structures is self-referential,
in that the structures and their morphisms contain (are indexed by) linguistic mechanisms for self-description.
This index is represented by a functor (passage between contexts)
$\mathrmbfit{lang} : \mathrmbf{Struc} \rightarrow \mathrmbf{Lang}$
from the category of semantical structures
to a category $\mathrmbf{Lang}$ of (logical) languages and language morphisms.
For any structure $M$,
$\mathrmbfit{lang}(M)$ is a language capable of describing $M$,
and for any structure morphism $f : M_1 \rightarrow M_2$,
$\mathrmbfit{lang}(f) : \mathrmbfit{lang}(M_1) \rightarrow \mathrmbfit{lang}(M_2)$ 
is a language morphism capable of describing $f$.
Hence,
a world consists of a triple
$\mathrmbf{Wrld} = \langle \mathrmbf{Struc}, \mathrmbf{Lang}, \mathrmbfit{lang} \rangle$ 
as discussed above.

The institutional approach assumes that multiple worlds exist and are relatable. 
In general,
different worlds use different languages for description.
Worlds are related through world morphisms.
A world morphism
$\mathrmbf{wrld} = \langle \mathrmbfit{struc}, \mathrmbfit{lang} \rangle 
: \mathrmbf{Wrld}_1 \rightarrow \mathrmbf{Wrld}_2$
consists of a structure functor $\mathrmbfit{struc} : \mathrmbf{Struc}_1 \rightarrow \mathrmbf{Struc}_2$
and a language functor $\mathrmbfit{lang} : \mathrmbf{Lang}_1 \rightarrow \mathrmbf{Lang}_2$
that commute with the source/target world linguistic indexings
$\mathrmbfit{struc} \circ \mathrmbfit{lang}_2 = \mathrmbfit{lang}_1 \circ \mathrmbfit{lang}$.
World morphisms are composable componentwise.
Let $\mathsfbf{World}$ denote the category consisting of worlds and world morphisms.


\begin{principle}{Polarity}
The linguistic indexing of the world is polar,
having equivalent homogeneous and heterogeneous forms.
\end{principle}

\noindent
We assume the linguistic indexing functor
$\mathrmbfit{lang} : \mathrmbf{Struc} \rightarrow \mathrmbf{Lang}$
is a fibration (Cartesian passage).
This gives a homogeneous representation of a world 
(all stuctures are in one category).
By the equivalence between fibrations and index categories,
we can alternately represent the world as the structure indexed category
$\mathrmbfit{struc} : \mathrmbf{Lang}^{\mathrm{op}} \rightarrow \mathrmbf{Cat}$
with the category of languages as its indexing category.
This gives a heterogeneous representation of a world 
(stuctures are in many categories, each category indexed by the language of structures in that category).
For any language $\Sigma$,
$\mathrmbfit{struc}(\Sigma)$ is the category
whose objects are structures with underlying language $\Sigma$,
and whose morphisms are structure morphisms whose underlying language morphism is the identity $1_\Sigma$.
Here the indexed category $\mathrmbfit{struc}$ is the heterogenization of the fibration $\mathrmbfit{lang}$,
and $\mathrmbfit{lang}$ is (up to equivalence) the homogenization (fusion) of $\mathrmbfit{struc}$.
The indexed category of stuctures models structural heterogeneity,
whereas 
the fibration of stuctures models structural homogeneity.
In summary,
a world can be represented as an indexed category
$\mathrmbf{Wrld} = \langle \mathrmbf{Lang}, \mathrmbfit{struc} 
: \mathrmbf{Lang}^{\mathrm{op}} \rightarrow \mathrmbf{Cat} \rangle$.

By the equivalence between fibration morphisms and index morphisms,
we can alternately represent a world morphism as an indexed morphism
$\langle \mathrmbfit{lang}, \mathrmbf{struc} \rangle 
: \langle \mathrmbf{Lang}_1, \mathrmbfit{struc}_1 \rangle 
  \rightarrow 
  \langle \mathrmbf{Lang}_2, \mathrmbfit{struc}_2 \rangle$,
where $\mathrmbfit{lang} : \mathrmbf{Lang}_1 \rightarrow \mathrmbf{Lang}_2$ is the language functor 
and 
$\mathrmbf{struc} 
: \mathrmbfit{struc}_1 \Rightarrow \mathrmbfit{lang}^{\mathrm{op}} \circ \mathrmbfit{struc}_2 
: \mathrmbf{Lang}_1^{\mathrm{op}} \rightarrow \mathrmbf{Cat}$ 
is a structure natural transformation (bridge between passages)
with $\Sigma^{\mathrm{th}}$ component functor 
$\mathrmbf{struc}_{\Sigma} 
: \mathrmbfit{struc}_1(\Sigma) \rightarrow \mathrmbfit{struc}_2(\mathrmbfit{lang}(\Sigma))$
for each language $\Sigma$ in $\mathrmbf{Lang}_1$,
where the naturality diagram 
$\mathrmbfit{struc}_1(\sigma) \circ \mathrmbf{struc}_{\Sigma} 
\cong 
\mathrmbf{struc}_{\Sigma^\prime} \circ \mathrmbfit{struc}_2(\mathrmbfit{lang}(\sigma))$
connecting the source/target structure indexes,
holds up to isomorphism
for each language morphism $\sigma : \Sigma \rightarrow \Sigma^\prime$ in $\mathrmbf{Lang}_1$.

The structure indexed category has an underlying indexed set
$|\mathrmbfit{struc}| = \mathrmbfit{struc} \circ |\mbox{-}| : \mathrmbf{Lang}^{\mathrm{op}} \rightarrow \mathrmbf{Set}$
obtained by forgetting morphism information;
any language $\Sigma$ is mapped to the set $|\mathrmbfit{struc}|(\Sigma)$ of all $\Sigma$-structures and any language morphism $\sigma : \Sigma_1 \rightarrow \Sigma_2$ is mapped to the structure reduct
$|\mathrmbfit{struc}|(\sigma) : |\mathrmbfit{struc}|(\Sigma_2) \rightarrow |\mathrmbfit{struc}|(\Sigma_1)$.

\subsection{Logical Expression}

\begin{principle}{Logic}
The description of the world involves semantics. 
It is logically meaningful, being based upon satisfaction.
\end{principle}

\noindent
We assume (syntax) that any structure can be described using logical expressions (sentences) built from its language. 
In particular,
we assume that there is a sentence functor (dual indexed set)
$\mathrmbfit{sen} : \mathrmbf{Lang} \rightarrow \mathrmbf{Set}$
with the category of languages as its source (indexing category);
any language $\Sigma$ is mapped to the set $\mathrmbfit{sen}(\Sigma)$ of all sentences built upon it
and any language morphism $\sigma : \Sigma_1 \rightarrow \Sigma_2$ is mapped to the sentence translation map 
$\mathrmbfit{sen}(\sigma) : \mathrmbfit{sen}(\Sigma_1) \rightarrow \mathrmbfit{sen}(\Sigma_2)$ built upon it.
Furthermore,
we assume (semantics) that the indexed sets of structures and sentences are linked by satisfaction;
any language $\Sigma$ is mapped to a satisfaction relation (truth classification)
${\models}_{\Sigma} \subseteq |\mathrmbfit{struc}|(\Sigma){\times}\mathrmbfit{sen}(\Sigma)$,
where
the symbolism $M \models_{\Sigma} s$ 
expressing the assertion that `$M$ satisfies $s$'
means that $\Sigma$-sentence $M$ is true when interpreted in $\Sigma$-structure $M$. 
A $\Sigma$-sentence $s \in \mathrmbfit{sen}(\Sigma)$ is a constraint (or theorem) of the $\Sigma$-structure $M$,
and is said to be valid in $M$, 
when $M$ satisfies $s$.

We also assume that satisfaction is preserved under sentence translation (infomorphism condition):
$|\mathrmbfit{struc}|(\Sigma)(M_2) \models_{\Sigma_1} s_1$
\underline{iff} 
$M_2 \models_{\Sigma_2} \mathrmbfit{sen}(\Sigma)(s_1)$
for target structure $M_2 \in |\mathrmbfit{struc}|(\Sigma_2)$ 
and source sentence  $s_1 \in \mathrmbfit{sen}(\Sigma_1)$. 
This expresses the invariance of truth under change of notation. 
Finally,
the structure and sentence indexed sets
can be combined with the satisfaction relations
into a functor 
$\mathrmbfit{cls} : \mathrmbf{Lang} \rightarrow \mathrmbf{Cls}$
from the category of languages to the category of classifications and infomorphisms \cite{kent:dagstuhl},
whose composition with instance/type component functors gives the structure/sentence indexed sets
$\mathrmbfit{cls}^{\mathrm{op}} \circ \mathrmbfit{inst} = |\mathrmbfit{struc}|$
and
$\mathrmbfit{cls} \circ \mathrmbfit{typ} = \mathrmbfit{sen}$.
We can think of the classification functor
$\mathrmbfit{cls} : \mathrmbf{Lang} \rightarrow \mathrmbf{Cls}$
as a diagram within the ambient category of classifications and infomorphisms,
indexed by languages and language morphisms.
When composed with the lift functor (see below)
logical expression and semantics extends to theories
$\mathrmbfit{cls} \circ \mathrmbfit{lift} : \mathrmbf{Lang} \rightarrow \mathrmbf{Cls}$.
	
\begin{principle}{Satisfaction}
The meaning of the world crucially involves both types and their particulars 
in classifications and infomorphisms.
\emph{(This is the transfer to satisfaction of the second principle of Information Flow \cite{barwise:seligman:97}.)}
\end{principle}
 
\noindent
This principle motivates the use of 
structures (via satisfaction) as the interpretative objects 
(for the local logics 
that incorporate the regularities of a distributed system)
and the use of structure morphisms (via truth invariance) as the interpretative morphisms 
(for the morphisms of local logics
that incorporate the information flow of regularities of a distributed system).

\subsection{Core Heterogeneity}

\begin{principle}{Core}
The architecture of the world description is concentrated in 
a 2-dimensional diagram of core indexed categories.
\end{principle}

\noindent
Being a classification relation,
satisfaction induces order.
The intent of a structure $M \in |\mathrmbfit{struc}|(\Sigma)$ 
is the theory
$M^\Sigma = \{ s \in \mathrmbfit{sen}(\Sigma) \mid M \models_\Sigma s \}$,
the set of all sentences that are satisfied by $M$;
that is,
the set of all theorems of $M$.
This is called the (maximal) theory of the structure $M$ and denoted be 
$\mathrmbfit{max}(M) = M^\Sigma$.
Structures are ordered by intent:
two structures $M_1,M_2 \in |\mathrmbfit{struc}|(\Sigma)$ 
are ordered $M_1 \leq_\Sigma M_2$ when $M_1^\Sigma \supseteq M_2^\Sigma$;
that is,
when $M_2 \models_\Sigma s$ implies $M_1 \models_\Sigma s$
for all sentences $s \in \mathrmbfit{sen}(\Sigma)$. 
This implies the condition,
if $M_1 \leq_\Sigma M_2$ and $M_2 \models_\Sigma s$ then $M_1 \models_\Sigma s$,
a bimodular condition 
stating that satisfaction respects the order on structures. 
%
Since preservation of satisfaction forms an infomorphism out of the structure and sentence maps,
the structure map lifts to a monotonic function:
$\mathrmbfit{struc}(\sigma) 
: \langle \mathrmbfit{struc}(\Sigma_2), \leq_{\Sigma_2} \rangle 
  \rightarrow 
  \langle \mathrmbfit{struc}(\Sigma_1), \leq_{\Sigma_1} \rangle$
for any language morphism $\sigma : \Sigma_1 \rightarrow \Sigma_2$.
This means that
the structure functor lifts to an indexed preorder (hence, category)
$\mathrmbfit{struc}^{\hat{\flat}} : \mathrmbf{Lang}^{\mathrm{op}} \rightarrow \mathrmbf{Pre} \subseteq \mathrmbf{Cat}$
with $\mathrmbfit{struc}^{\hat{\flat}} \circ |\,\mbox{-}\,| = |\mathrmbfit{struc}|$.

For any language $\Sigma$,
a theory is a subset of sentences $T \in {\wp}\mathrmbfit{sen}(\Sigma)$. 
Satisfaction lifts to theories:
a structure satisfies a theory $M \models_\Sigma T$ when it satisfies every sentence in it;
then $T$ is said to be valid in $M$.
A theory $T$ entails a sentence $s$,
$T \vdash_\Sigma s$,
when any structure that satisfies $T$ also satisfies $s$.
The extent of a theory 
$T^\Sigma$
is the set
of all structures that satisfy it.
The closure of a theory is the set of all entailed sentences 
--- the intent of the extent $T^\bullet = T^{\Sigma\Sigma}$.
Theories are ordered (entailment) by extent:
two theories $T_1,T_2 \in {\wp}\mathrmbfit{sen}(\Sigma)$ 
are ordered $T_1 \leq_\Sigma T_2$ when $T_1^\Sigma \subseteq T_2^\Sigma$.
This implies the bimodular condition,
if $M \models_\Sigma T_1$ and $T_1 \leq_\Sigma T_2$ then $M \models_\Sigma T_2$,
which states that satisfaction respects entailment order. 
The sentence functor extends to theories in two adjoint ways.
There is a direct image functor (dual indexed preorder)
$\mathrmbfit{dir} : \mathrmbf{Lang} \rightarrow \mathrmbf{Pre}$;
any language $\Sigma$ is mapped to the entailment preorder 
$\mathrmbfit{th}(\Sigma) 
= \langle {\wp}\mathrmbfit{sen}(\Sigma), \leq_{\Sigma} \rangle$ 
of all theories built upon it, 
and any language morphism $\sigma : \Sigma_1 \rightarrow \Sigma_2$ 
is mapped to the theory direct image monotonic function 
$\mathrmbfit{dir}(\sigma) 
: \mathrmbfit{th}(\Sigma_1) \rightarrow \mathrmbfit{th}(\Sigma_2)$;
there is a related functor $\mathrmbfit{lift} : \mathrmbf{Cls} \rightarrow \mathrmbf{Cls}$
that lifts the direct image functor on $\mathrmbf{Set}$.
There is an inverse image functor (indexed preorder)
$\mathrmbfit{inv} : \mathrmbf{Lang}^{\mathrm{op}} \rightarrow \mathrmbf{Pre}$;
any language $\Sigma$ is mapped to the entailment preorder 
$\mathrmbfit{th}(\Sigma)$, 
and any language morphism $\sigma : \Sigma_1 \rightarrow \Sigma_2$ 
is mapped to the theory inverse image monotonic function 
$\mathrmbfit{inv}(\sigma) 
: \mathrmbfit{th}(\Sigma_2) \rightarrow \mathrmbfit{th}(\Sigma_1)$.

Structures and theories can be embedded monotonically as concepts in the concept lattice of satisfaction \cite{ganter:wille:99}. 
Due to these embeddings,
the intent order on structures and the entailment order on theories 
are special cases of the lattice order of the satisfaction concept lattice.
The concept lattice order is a generalization-specialization order
with the more general concepts above and the more special concepts below.
The fiber order of stuctures (theories) 
is the opposite of 
the intent (entailment) order on structures (theories) 
induced by the satisfaction concept lattice.
Composition with the opposite preorder involution ${\propto} : \mathrmbf{Pre} \rightarrow \mathrmbf{Pre}$
gives 
the structure involuted indexed preorder
$\mathrmbfit{struc}^{\flat} = \mathrmbfit{struc}^{\hat{\flat}} \circ {\propto} 
: \mathrmbf{Lang}^{\mathrm{op}} \rightarrow \mathrmbf{Pre}$
and
the theory inverse image involuted indexed preorder
$\mathrmbfit{inv}^{\propto} = \mathrmbfit{inv} \circ {\propto} 
: \mathrmbf{Lang}^{\mathrm{op}} \rightarrow \mathrmbf{Pre}$.
The structure involuted indexed preorder gives a heterogeneous representation of a flat world.

\subsection{Flat Structures}

Associated (homogenization) with the structure involuted indexed preorder $\mathrmbfit{struc}^{\flat}$
is the (flattened) structure fibration
$\mathrmbfit{lang}^{\scriptscriptstyle\flat} 
: \mathrmbf{Struc}^{\scriptscriptstyle\flat} \rightarrow \mathrmbf{Lang}$
define as follows.
$\mathrmbf{Struc}^{\scriptscriptstyle\flat}$,
the category of (flat) structures and (flat) structure morphisms,
is the Grothendieck construction of $\mathrmbfit{struc}^{\flat}$.
A (flat) structure $\langle \Sigma, M \rangle$ consists of 
a language $\Sigma$ and a $\Sigma$-structure $M \in \mathrmbfit{struc}(\Sigma)$
(so that $\mathrmbfit{lang}(M) = \Sigma$).
A (flat) structure morphism
$\sigma : \langle \Sigma_1, M_1 \rangle \rightarrow \langle \Sigma_2, M_2 \rangle$
is a language morphism $\sigma : \Sigma_1 \rightarrow \Sigma_2$
that preserves constraints (theorems):
$M_1 \models_{\Sigma_1} s_1$ implies $M_2 \models_{\Sigma_2} \mathrmbfit{sen}(\sigma)(s_1)$. 
Equivalently,
a (flat) structure morphism is an language morphism 
whose structure component maps the target structure to a specialization of the source structure
$\mathrmbfit{struc}(\sigma)(M_2) \leq_{\Sigma_1} M_1$
or (intentwise, using theories)
$\mathrmbfit{struc}(\sigma)(M_2)^{\Sigma_1} \leq_{\Sigma_1} M_1^{\Sigma_1}$
Equivalently,
a (flat) structure morphism is an language morphism 
whose inverse image sentence component maps the target structure intent to a specialization of the source structure intent
$\mathrmbfit{sen}(\sigma)^{-1}(M_2^{\Sigma_2}) \leq_{\Sigma_1} M_1^{\Sigma_1}$.
This gives a homogeneous representation of a flat world,
the fiber flattening of the category $\mathrmbf{Struc}$.
The (flattened) language functor $\mathrmbfit{lang}^{\scriptscriptstyle\flat}$
is the projection,
which maps an object $\langle \Sigma, M \rangle$
to its indexing language $\Sigma$ 
and maps a morphism $\sigma : \langle \Sigma_1, M_1 \rangle \rightarrow \langle \Sigma_2, M_2 \rangle$
to its indexing language morphism $\sigma : \Sigma_1 \rightarrow \Sigma_2$.
This is a fibration.
For any language $\Sigma$,
the identity language morphism $1_\Sigma : \Sigma \rightarrow \Sigma$
is a (flat) structure morphism
$1_\Sigma : \langle \Sigma, M_1 \rangle \rightarrow \langle \Sigma, M_2 \rangle$ \underline{iff} $M_1 \geq_\Sigma M_2$.


\begin{principle}{Bimodular}
The description of the world factors through flat structures.
Satisfaction is bimodular;
that is,
satisfaction respects structure morphisms.
\end{principle}

\noindent
We assume that the language fibration factors 
$\mathrmbfit{lang} = \mathrmbfit{flat} \circ \mathrmbfit{lang}^{\flat}$
through the flattened language fibration 
$\mathrmbfit{lang}^{\flat} : \mathrmbf{Struc}^{\flat} \rightarrow \mathrmbf{Lang}$
by way of a structure flattening functor
$\mathrmbfit{flat} : \mathrmbf{Struc} \rightarrow \mathrmbf{Struc}^{\flat}$.
This means the following:
For any structure $M$,
if $\mathrmbfit{lang}(M) = \Sigma$
then $\mathrmbfit{flat}(M) = \langle \Sigma, M \rangle$.
For any structure morphism $f : M_1 \rightarrow M_2$,
if $\mathrmbfit{lang}(f : M_1 \rightarrow M_2) = \sigma : \Sigma_1 \rightarrow \Sigma_2$,
then
$\mathrmbfit{flat}(f : M_1 \rightarrow M_2)
= \sigma : \langle \Sigma_1, M_1 \rangle \rightarrow \langle \Sigma_2, M_2 \rangle$.
Hence,
$M_1 \geq_{\Sigma_1} \mathrmbfit{struc}(\sigma)(M_2)$;
equivalently, 
$M_1 \models_{\Sigma_1} s_1$ implies 
$M_2 \models_{\Sigma_2} \mathrmbfit{sen}(\sigma)(s_1)$
(or $\mathrmbfit{struc}(\sigma)(M_2) \models_{\Sigma_1} s_1$)
for every sentence $s_1 \in \mathrmbfit{sen}(\Sigma_1)$.
This implies the condition,
if $f : M_1 \rightarrow M_2$ is a structure morphism and $M_1 \models_{\Sigma_1} s_1$
then $M_2 \models_{\Sigma_2} \mathrmbfit{sen}(\sigma)(s_1)$,
a bimodular condition 
stating that satisfaction respects structure morphisms. 
In particular,
if $f : M_1 \rightarrow M_2$ is a vertical structure morphism over language $\Sigma$,
then
$M_1 \models_{\Sigma} s$ implies $M_2 \models_{\Sigma} s$ for any sentence $s \in \mathrmbfit{sen}(\Sigma)$; 
that is, $M_1 \geq_\Sigma M_2$.
Hence,
the (flat) structure fiber over $\Sigma$ is the underlying preorder of the structure fiber.


\subsection{Theories}

Associated (homogenization) with the theory inverse image involuted indexed preorder $\mathrmbfit{inv}^{\propto}$
is the theory fibration
$\mathrmbfit{lang} : \mathrmbf{Th} \rightarrow \mathrmbf{Lang}$
define as follows.
$\mathrmbf{Th}$, 
the category of theories and theory morphisms,
is the Grothendieck construction of $\mathrmbfit{inv}^{\propto}$.
A theory $\langle \Sigma, T \rangle$ consists of 
a language $\Sigma$ and a $\Sigma$-theory $T \in {\wp}\mathrmbfit{sen}(\Sigma)$.
A theory morphism
$\sigma : \langle \Sigma_1, T_1 \rangle \rightarrow \langle \Sigma_2, T_2 \rangle$
is a language morphism $\sigma : \Sigma_1 \rightarrow \Sigma_2$
that maps the target theory to a specialization of the source theory
$\mathrmbfit{inv}(\sigma)(T_2) = \mathrmbfit{sen}(\sigma)^{-1}(T_2^\bullet) \leq_{\Sigma_1} T_1$
\underline{iff} $\mathrmbfit{sen}(\sigma)^{-1}(T_2^\bullet) \supseteq T_1$
\underline{iff} $\mathrmbfit{sen}(\sigma)^{-1}(T_2^\bullet) \supseteq T_1^\bullet$;
or that preserves entailment,
$T_1 \vdash_{\Sigma_1} t_1$
implies 
$T_2 \vdash_{\Sigma_2} \mathrmbfit{sen}(\sigma)(t_1)$
for any $t_1 \in \mathrmbfit{sen}(\Sigma_1)$.
Equivalently,
a theory morphism is an language morphism 
that maps the source theory to a generalization of the target theory
$T_2 \leq_{\Sigma_2} \mathrmbfit{dir}(\sigma)(T_1)
= {\wp}\mathrmbfit{sen}(\sigma)(T_1)$
\underline{iff} 
$T_2^\bullet \supseteq {\wp}\mathrmbfit{sen}(\sigma)(T_1)$.
The projection fibration
$\mathrmbfit{lang} : \mathrmbf{Th} \rightarrow \mathrmbf{Cls}$
maps a theory $\langle \Sigma, T \rangle$ to the language $\Sigma$ 
and maps a theory morphism
$\sigma : \langle \Sigma_1, T_1 \rangle \rightarrow \langle \Sigma_2, T_2 \rangle$
to the language morphism $\sigma : \Sigma_1 \rightarrow \Sigma_2$.
For any language $\Sigma$,
the identity language morphism $1_\Sigma : \Sigma \rightarrow \Sigma$
is a (vertical) theory morphism
$1_\Sigma : \langle \Sigma, T_1 \rangle \rightarrow \langle \Sigma, T_2 \rangle$ \underline{iff} $T_1 \geq_A T_2$.
Hence,
the theory fiber at language $\Sigma$
is the opposite of the entailment theory preorder.

\subsection{Logical Environments}

\begin{figure}
\begin{center}
\begin{tabular}[b]{c@{\hspace{80pt}}c}
 & \\ & \\
\setlength{\unitlength}{0.8pt}
\begin{picture}(100,120)(-20,0)
\put(-20,120){\makebox(0,0){$\mathrmbf{Log}$}}
\put(-60,60){\makebox(0,0){$\mathrmbf{Struc}$}}
\put(40,120){\makebox(0,0){$\mathrmbf{Log}^{\scriptscriptstyle\flat}$}}
\put(5,60){\makebox(0,0){$\mathrmbf{Struc}^{\scriptscriptstyle\flat}$}}
\put(40,0){\makebox(0,0){$\mathrmbf{Lang}$}}
\put(80,60){\makebox(0,0){$\mathrmbf{Th}$}}
\put(-37,18){\makebox(0,0)[r]{\footnotesize{$\mathrmbfit{lang}$}}}
\put(17,127){\makebox(0,0)[r]{\scriptsize{$\mathrmbfit{flat}$}}}
\put(-23,67){\makebox(0,0)[r]{\scriptsize{$\mathrmbfit{flat}$}}}
\put(15,30){\makebox(0,0)[r]{\footnotesize{$\mathrmbfit{lang}^{\scriptscriptstyle\flat}$}}}
\put(40,67){\makebox(0,0){\footnotesize{$\mathrmbfit{max}$}}}
\put(67,28){\makebox(0,0)[l]{\footnotesize{$\mathrmbfit{lang}$}}}
\put(-44,93){\makebox(0,0)[r]{\scriptsize{$\mathrmbfit{pr}_0$}}}
\put(5,90){\makebox(0,0)[r]{\scriptsize{$\mathrmbfit{pr}_1$}}}
\put(42,93){\makebox(0,0)[r]{\scriptsize{$\mathrmbfit{pr}_0$}}}
\put(82,93){\makebox(0,0)[r]{\scriptsize{$\mathrmbfit{pr}_1$}}}
\qbezier(40,37)(45,32)(50,27)
\qbezier(40,37)(35,32)(30,27)
\qbezier(30,37)(32,35)(34,33)
\qbezier(36,31)(38,29)(45,22)
\qbezier(30,37)(27,34)(22,29)
\qbezier(20,27)(19,26)(10,17)
\put(-30,108){\vector(-2,-3){22}}
\put(-2,120){\vector(1,0){22}}
\put(-40,60){\vector(1,0){18}}
\qbezier(-58,43)(-25,10)(12,0)
\put(12,0){\vector(4,-1){0}}
\qbezier(-13,108)(15,82)(60,70)
\put(60,70){\vector(4,-1){0}}
\put(30,108){\vector(-2,-3){22}}
\put(50,108){\vector(2,-3){22}}
\put(25,60){\vector(1,0){35}}
\put(8,48){\vector(2,-3){22}}
\put(72,48){\vector(-2,-3){22}}
\put(40,-55){\makebox(0,0){\footnotesize{\emph{fibration}}}}
\end{picture}
&
\setlength{\unitlength}{0.8pt}
\begin{picture}(80,120)(0,0)
\put(0,75){\begin{picture}(80,60)(0,0)
\put(80,60){\makebox(0,0){$\mathrmbf{Set}$}}
\put(-3,60){\makebox(0,0){$\mathrmbf{Cat}$}}
\put(-3,0){\makebox(0,0){$\mathrmbf{Lang}$}}
\put(80,0){\makebox(0,0){$\mathrmbf{Cls}$}}
\put(40,10){\makebox(0,0){\footnotesize{$\mathrmbfit{cls}$}}}
\put(-16,28){\makebox(0,0){\footnotesize{$\mathrmbfit{sen}$}}}
\put(28,35){\makebox(0,0){\footnotesize{$\mathrmbfit{sen}$}}}
\put(87,28){\makebox(0,0)[l]{\footnotesize{$\mathrmbfit{typ}$}}}
\put(40,70){\makebox(0,0){\scriptsize{$|\,\mbox{-}\,|$}}}
\put(20,0){\vector(1,0){40}}
\put(20,60){\vector(1,0){40}}
\put(-3,12){\vector(0,1){36}}
\put(80,12){\vector(0,1){36}}
\put(16,12){\vector(4,3){48}}
\end{picture}}
\put(0,-15){\begin{picture}(80,60)(0,0)
\put(-3,60){\makebox(0,0){$\mathrmbf{Lang}^{\mathrm{op}}$}}
\put(80,60){\makebox(0,0){$\mathrmbf{Cls}^{\mathrm{op}}$}}
\put(-3,0){\makebox(0,0){$\mathrmbf{Cat}$}}
\put(80,0){\makebox(0,0){$\mathrmbf{Set}$}}
\put(-20,30){\makebox(0,0)[r]{\footnotesize{$\mathrmbfit{struc}$}}}
\put(10,31){\makebox(0,0)[l]{\footnotesize{$\mathrmbfit{struc}^{\flat}$}}}
\put(-5,35){\makebox(0,0){\tiny{$\mathrmbf{struc}$}}}
\put(-5,27){\makebox(0,0){\footnotesize{$\Longrightarrow$}}}
\put(40,70){\makebox(0,0){\footnotesize{$\mathrmbfit{cls}^{\mathrm{op}}$}}}
\put(87,32){\makebox(0,0)[l]{\footnotesize{$\mathrmbfit{inst}$}}}
\put(64,42){\makebox(0,0)[r]{\footnotesize{$|\mathrmbfit{struc}|$}}}
\put(40,-10){\makebox(0,0){\scriptsize{$|\,\mbox{-}\,|$}}}
\put(16,48){\line(4,-3){16}}
\put(42,28.5){\vector(4,-3){24}}
\qbezier(-13,48)(-23,30)(-13,12)
\put(-13,12){\vector(1,-3){0}}
\qbezier(2,48)(12,30)(2,12)
\put(2,12){\vector(-1,-3){0}}
\put(80,48){\vector(0,-1){36}}
\put(20,60){\vector(1,0){40}}
\put(20,0){\vector(1,0){40}}
\end{picture}}
\put(40,-55){\makebox(0,0){\footnotesize{\emph{indexed category}}}}
\end{picture}
\\ & \\ & \\ & \\
\multicolumn{2}{c}{
\begin{picture}(0,0)(0,0)
\put(0,12){\makebox(0,0){\scriptsize{\emph{heterogenization}}}}
\put(-30,5){\vector(1,0){60}}
\put(30,-5){\vector(-1,0){60}}
\put(0,-12){\makebox(0,0){\scriptsize{\emph{homogenization}}}}
\end{picture}} \\ 
& \\
\end{tabular}
\end{center}
\caption{Logical Environment}
\label{logical:environment}
\end{figure}
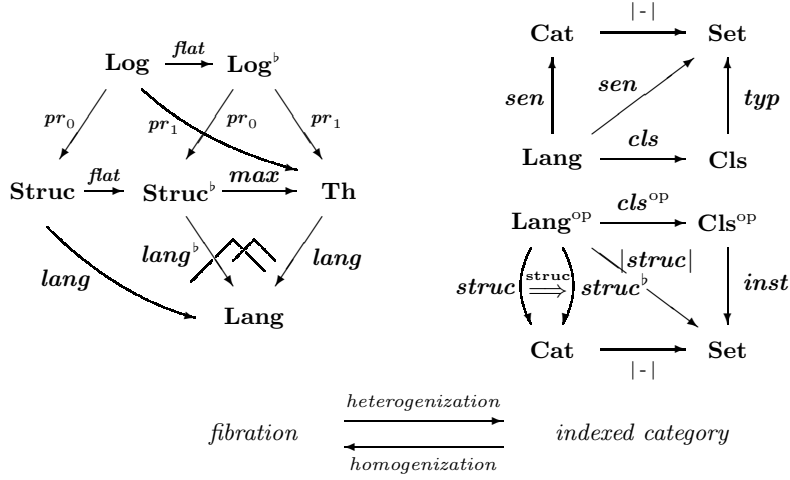

\noindent
In summary,
a \emph{logical environment} (Figure~\ref{logical:environment}) is a more semantic version of an institution.
It has both semantical and logical aspects.
The semantical aspect is represented by
a category of structures $\mathrmbf{Struc}$ (the world) with a fibration 
$\mathrmbfit{lang} : \mathrmbf{Struc} \rightarrow \mathrmbf{Lang}$ (the world description)
from structures into a category of logical languages $\mathrmbf{Lang}$;
or equivalently,
an indexed category of structures
$\mathrmbfit{struc} : \mathrmbf{Lang}^{\mathrm{op}} \rightarrow \mathrmbf{Cat}$
with underlying indexed set
$|\mathrmbfit{struc}| 
= \mathrmbfit{struc} \circ |\mbox{-}| : \mathrmbf{Lang}^{\mathrm{op}} \rightarrow \mathrmbf{Set}$.
The logical aspect is represented by
a dual indexed set of sentences
$\mathrmbfit{sen} : \mathrmbf{Lang} \rightarrow \mathrmbf{Set}$.
The semantical aspect is connected to the logical aspect via a functor 
$\mathrmbfit{cls} : \mathrmbf{Lang} \rightarrow \mathrmbf{Cls}$,
with instance projection 
$\mathrmbfit{cls}^{\mathrm{op}} \circ \mathrmbfit{inst} = |\mathrmbfit{struc}|$
and type projection 
$\mathrmbfit{cls} \circ \mathrmbfit{typ} = \mathrmbfit{sen}$.
Structures and sentences are linked by satisfaction $M \models_{\Sigma} s$,
with truth preserved under change of notation 
$\mathrmbfit{struc}(M_2) \models_{\Sigma_1} s_1$ \underline{iff} $M_2 \models_{\Sigma_1} \mathrmbfit{sen}(s_1)$.
The concept order of satisfaction 
extentionally lifts to the (flat) structure indexed preorder 
$\mathrmbfit{struc}^{\flat} : \mathrmbf{Lang}^{\mathrm{op}} \rightarrow \mathrmbf{Pre}$
and 
intentionally lifts to the theory direct image dual indexed preorder
$\mathrmbfit{dir} : \mathrmbf{Lang} \rightarrow \mathrmbf{Pre}$
and (adjointly)
to the theory inverse image indexed preorder
$\mathrmbfit{inv} : \mathrmbf{Lang}^{\mathrm{op}} \rightarrow \mathrmbf{Pre}$.
The (flat) structure indexed preorder
is equivalent (homogenization $\leftrightarrow$ heterogenization) to 
the (flattened) structure fibration (the flattened world) 
$\mathrmbfit{lang}^{\flat} : \mathrmbf{Struc}^{\flat} \rightarrow \mathrmbf{Lang}$,
such that description factors through flat structures
$\mathrmbfit{lang} = \mathrmbfit{flat} \circ \mathrmbfit{lang}^{\flat}$. 
Either the theory direct image dual indexed preorder 
or the theory inverse image indexed preorder 
induce the theory fibration (the logical aspect)
$\mathrmbfit{lang} : \mathrmbf{Th} \rightarrow \mathrmbf{Lang}$.
The crucial facts linking the semantical aspect of structures to the logical aspect of theories via satisfaction
are (1) that the indexed set of structures is the instance projection of the classification functor
and (2) that description factors through flat structures. 

\subsection{Examples}

We use the three logical environments of 
equational logic $\mathtt{EQ}$, first order logic $\mathtt{FOL}$ and information flow $\mathtt{IFC}$
as running examples.
The world (structure category) of $\mathtt{EQ}$ consists of universal algebriac structures,
that of $\mathtt{FOL}$ consists of first order logical structures,
and that of $\mathtt{IFC}$ consists of classifications and infomorphisms. 

\paragraph{Example.} 
The logical environment $\mathttbf{IFC}$ has 
$\mathrmbf{Struc}_{\scriptscriptstyle\mathttbf{IFC}} = \mathrmbf{Cls}$ as its category of structures,
$\mathrmbf{Lang}_{\scriptscriptstyle\mathttbf{IFC}} = \mathrmbf{Set}$ as its category of languages,
and the type functor
$\mathrmbfit{lang}_{\scriptscriptstyle\mathttbf{IFC}} = \mathrmbfit{typ} 
: \mathrmbf{Cls} \rightarrow \mathrmbf{Set}$ as its projection.
For any set $Y$,
the set $\mathrmbfit{struc}_{\scriptscriptstyle\mathttbf{IFC}}(Y)$ of $Y$-structures 
is the set of classifications with type set $Y$,
the set $\mathrmbfit{sen}_{\scriptscriptstyle\mathttbf{IFC}}(Y)$ of $Y$-sentences 
is the set of $Y$-sequents,
pairs $\mathrmit{\Gamma} \vdash \mathrmit{\Delta}$ 
of subsets of types $\mathrmit{\Gamma},\mathrmit{\Delta} \subseteq Y$,
and 
a $Y$-classication $\langle X, Y, \models \rangle$ 
satisfies a $Y$-sequent $\mathrmit{\Gamma} \vdash \mathrmit{\Delta}$,
denoted $\langle X, Y, \models \rangle \models_{Y} (\mathrmit{\Gamma} \vdash \mathrmit{\Delta})$,
when for all instances $x \in X$, 
$x \models y$ for all $y \in \mathrmit{\Gamma}$ implies $x \models y^\prime$ for some $y^\prime \in \mathrmit{\Delta}$.
For each function (language translation) $f : Y \rightarrow Z$,
sentence translation along $f$ is direct image squared on types
$\mathrmbfit{sen}_{\scriptscriptstyle\mathttbf{IFC}}(f)
:             \mathrmbfit{sen}_{\scriptscriptstyle\mathttbf{IFC}}(Y) 
  \rightarrow \mathrmbfit{sen}_{\scriptscriptstyle\mathttbf{IFC}}(Z)
: (\mathrmit{\Gamma} \vdash \mathrmit{\Delta}) 
  \mapsto
  ({\wp}f(\mathrmit{\Gamma}) \vdash {\wp}f(\mathrmit{\Delta}))$,
and structure translation along $f$, 
$\mathrmbfit{struc}_{\scriptscriptstyle\mathttbf{IFC}}(f)
:             \mathrmbfit{struc}_{\scriptscriptstyle\mathttbf{IFC}}(Z) 
  \rightarrow \mathrmbfit{struc}_{\scriptscriptstyle\mathttbf{IFC}}(Y)$,
maps a $Z$-classification $C = \langle X, Z, \models \rangle$ to the $Y$-classification 
$\mathrmbfit{struc}_{\scriptscriptstyle\mathttbf{IFC}}(f)(C) = \langle X, Y, \models_f \rangle$,
where $x \models_f y$ when $x \models f(y)$.
The classification functor 
$\mathrmbfit{cls}_{\scriptscriptstyle\mathttbf{IFC}} : \mathrmbf{Set} \rightarrow \mathrmbf{Cls}$
maps a set $Y$ to the classification 
$\mathrmbfit{cls}_{\scriptscriptstyle\mathttbf{IFC}}(Y) 
= \langle \mathrmbfit{struc}_{\scriptscriptstyle\mathttbf{IFC}}(Y), 
          \mathrmbfit{sen}_{\scriptscriptstyle\mathttbf{IFC}}(Y), 
          \models_Y \rangle$,
and
maps a function $f : Y \rightarrow Z$ to the infomorphism
$\mathrmbfit{cls}_{\scriptscriptstyle\mathttbf{IFC}}(f) 
= \langle \mathrmbfit{struc}_{\scriptscriptstyle\mathttbf{IFC}}(f), 
          \mathrmbfit{sen}_{\scriptscriptstyle\mathttbf{IFC}}(f) \rangle 
          :        \mathrmbfit{cls}_{\scriptscriptstyle\mathttbf{IFC}}(Y) 
\rightleftharpoons \mathrmbfit{cls}_{\scriptscriptstyle\mathttbf{IFC}}(Z)$.
The logical environment $\mathttbf{IFC}$ is a subenvironment of $\mathttbf{FOL}$
when types are regarded as unary relation symbols. 

\paragraph{Example.} 
The logical system of equational logic (universal algebra) 
is represented by the logical environment {\ttfamily EQ}. 
The language category is 
$\mathrmbf{Lang}_{\scriptscriptstyle\mathttbf{EQ}} = \mathrmbf{Set}^{\aleph}$,
the $\aleph^{\mathrm{th}}$ power of $\mathrmbf{Set}$.
A language $\Phi$ is a family $\Phi = \{ \Phi_n \mid n \in \aleph \}$ of sets of function symbols, and 
a language morphism $\phi : \Phi \rightarrow \Phi^\prime$ 
is a family $\{ \phi_n : \Phi_n \rightarrow \Phi_n^\prime \mid n \in \aleph \}$ 
of arity-preserving maps of function symbols. 
For any language $\Phi$, 
the set $\mathrmbfit{sen}_{\scriptscriptstyle\mathtt{EQ}}(\Phi)$ is the set of equations between $\Phi$-terms of function symbols. 
For any language morphism $\phi : \Phi \rightarrow \Phi^\prime$, 
the sentence translation function 
$\mathrmbfit{sen}_{\scriptscriptstyle\mathtt{EQ}}(\phi) 
: \mathrmbfit{sen}_{\scriptscriptstyle\mathtt{EQ}}(\Phi) \rightarrow \mathrmbfit{sen}_{\scriptscriptstyle\mathtt{EQ}}(\Phi^\prime)$ 
is defined by function symbol substitution. 
A $\Phi$-structure $A \in \mathrmbfit{struc}_{\scriptscriptstyle\mathtt{EQ}}(\Phi)$ is a $\Phi$-algebra, 
consisting of a set (universe) $A$ 
and a function (operation) $f_A : A^n \rightarrow A$ for each function symbol $f \in \Phi_n$. 
A $\Phi$-structure morphism in $\mathrmbfit{struc}_{\scriptscriptstyle\mathtt{EQ}}(\Phi)$ is a $\Phi$-algebra morphism, 
consisting of a function (between universes) $a : A \rightarrow A^\prime$ 
that preserves operations $f_A \cdot a = a^n \cdot f_{A^\prime}$ for each function symbol $f \in \Phi_n$. 
Structure translation is reduct with symbol translation. Satisfaction is as usual. 

\paragraph{Example.} 
The logical system of unsorted first-order logic with equality 
is represented by the logical environment {\ttfamily FOL}. 
This extends the logical environment of equational logic by adding relation symbols. 
The language category is 
$\mathrmbf{Lang}_{\scriptscriptstyle\mathttbf{FOL}} = \mathrmbf{Set}^{\aleph}{\!\times}\mathrmbf{Set}^{\aleph}
\cong \mathrmbf{Set}^{\aleph+\aleph}
\cong \mathrmbf{Set}^{\aleph}$,
the square of the $\aleph^{\mathrm{th}}$ power of $\mathrmbf{Set}$.
A language $\langle \Phi, \Psi \rangle$ is a family $\Phi$ as above, 
plus a family $\Psi = \{ \Psi_n \mid n \in \aleph \}$ 
of sets of relation symbols of arity $n$. 
A language morphism 
$\langle \phi, \psi \rangle : \langle \Phi, \Psi \rangle \rightarrow \langle \Phi^\prime, \Psi^\prime \rangle$ 
is a family $\phi$ as above, 
plus a family $\psi : \Psi \rightarrow \Psi^\prime$ of arity-preserving maps of relation symbols. 
Sentences are the usual first order sentences. 
For any language $\langle \Phi, \Psi \rangle$, the set 
$\mathrmbfit{sen}_{\scriptscriptstyle\mathtt{FOL}}(\Phi, \Psi)$ of $\langle \Phi, \Psi \rangle$-sentences 
consists of closed first-order formulae using function symbols from $\Phi$ and relation symbols from $\Psi$. 
For any language morphism 
$\langle \phi, \psi \rangle : \langle \Phi, \Psi \rangle \rightarrow \langle \Phi^\prime, \Psi^\prime \rangle$, 
the sentence translation function 
$\mathrmbfit{sen}_{\scriptscriptstyle\mathtt{FOL}}(\phi, \psi) 
: \mathrmbfit{sen}_{\scriptscriptstyle\mathtt{FOL}}(\Phi, \Psi) \rightarrow \mathrmbfit{sen}_{\scriptscriptstyle\mathtt{FOL}}(\Phi^\prime, \Psi^\prime)$ 
is defined by symbol substitution. 
A $\langle\Phi,\Psi\rangle$-structure $A \in \mathrmbfit{struc}_{\scriptscriptstyle\mathtt{FOL}}(\Phi, \Psi)$ 
is a $\Phi$-algebra $A$ (as above)
and a subset $R_A \subseteq A^n$ for each relation symbol $R \in \Psi_n$. 
A $\langle\Phi,\Psi\rangle$-structure morphism in $\mathrmbfit{struc}_{\scriptscriptstyle\mathtt{EQ}}(\Phi)$ 
is a $\Phi$-algebra morphism (as above),
which preserves relations ${\wp}a^n(R_A) \subseteq R_{A^\prime}$ for each relation symbol $R \in \Psi_n$. 
Structure translation is reduct with symbol translation. 
Satisfaction is as usual.
The institution {\ttfamily FOL} can be extended to the institution $\mathtt{FOL}^\ast$, 
which replaces language maps with language interpretations 
$\langle \phi, \psi \rangle : \langle \Phi, \Psi \rangle \rightarrow \langle \Phi^\prime, \Psi^\prime \rangle$ 
mapping function symbols to terms of the same arity 
$\{ \phi_n : \Phi_n \rightarrow \mathrmbfit{term}(\Phi^\prime)_n \mid n \in \aleph \}$ 
and mapping relation symbols to expressions of the same arity 
$\{ \psi_n : \Psi_n \rightarrow \mathrmbfit{expr}(\Psi^\prime)_n \mid n \in \aleph \}$.

\paragraph{Example.} 
The category $\mathrmbf{Set}_{\scriptscriptstyle{\subseteq}}$ 
has subsets $Y \subseteq X$ as objects
and restrictions $(g \subseteq f) : (Y_1 \subseteq X_1) \rightarrow (Y_2 \subseteq X_2)$ as morphisms,
where $f : X_1 \rightarrow X_2$ is a function and $g : Y_1 \rightarrow Y_2$ is a restriction of $f$.
The category $\mathrmbf{Cls}_{\scriptscriptstyle{\subseteq}}$ 
has $\mathrmbf{Set}_{\scriptscriptstyle{\subseteq}}$ as its component instance category
and $\mathrmbf{Set}$ as its component type category.
An object 
$\langle Y, A \rangle$
in $\mathrmbf{Cls}_{\scriptscriptstyle{\subseteq}}$
consists of a classification $A$ and a subset of instances $Y \subseteq \mathrmbfit{inst}(A)$.
A morphism 
$\langle g, f \rangle : \langle Y_1, A_1 \rangle \rightleftharpoons \langle Y_2, A_2 \rangle$
in $\mathrmbf{Cls}_{\scriptscriptstyle{\subseteq}}$
consists of an infomorphism $f : A_1 \rightleftharpoons A_2$
and a restriction $g : Y_2 \rightarrow Y_1$ of the instance function $\mathrmbfit{inst}(f)$.
The logical environment $\mathttbf{IFS}$ 
has 
$\mathrmbf{Lang}_{\scriptscriptstyle\mathtt{IFS}}  = \mathrmbf{Set}$ as its category of languages,
$\mathrmbf{Struc}_{\scriptscriptstyle\mathtt{IFS}} = \mathrmbf{Cls}_{\scriptscriptstyle{\subseteq}}$ 
as its category of structures,
and
$\mathrmbfit{lang}_{\scriptscriptstyle\mathtt{IFS}} = \mathrmbfit{typ} 
: \mathrmbf{Cls}_{\scriptscriptstyle{\subseteq}} \rightarrow \mathrmbf{Set}$
as its language index functor.
This logical environment allows the definition of a normal subset of instances.

\section{Information Flow}


\subsection{Distributed Systems}


\begin{principle}{System}
Information flow results from regularities in a distributed system.
\emph{(This is the first principle of Information Flow \cite{barwise:seligman:97}.)}
\end{principle}

\noindent
This principle motivates the representation of distributed systems by diagrams 
of objects that can incorporate regularities.
Eventually,
we will argue that these objects should be local logics.

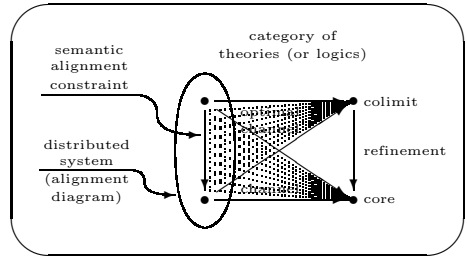
\begin{figure}
\begin{center}
\begin{tabular}{c}
\setlength{\unitlength}{0.95pt}
\begin{picture}(200,100)(0,0)
\put(25,85){\oval(30,30)[tl]}
\put(25,15){\oval(30,30)[bl]}
\put(175,85){\oval(30,30)[tr]}
\put(175,15){\oval(30,30)[br]}
\put(25,100){\line(1,0){150}}
\put(25,0){\line(1,0){150}}
\put(10,85){\line(0,-1){70}}
\put(190,85){\line(0,-1){70}}
\put(-3,-3){\begin{picture}(0,0)(0,0)
\put(125,90){\makebox(0,0){\tiny{category of}}}
\put(125,83){\makebox(0,0){\tiny{theories (or logics)}}}
\end{picture}}
\put(-2,2){\begin{picture}(0,0)(0,0)
\put(43,80){\makebox(0,0){\tiny{semantic}}}
\put(43,73){\makebox(0,0){\tiny{alignment}}}
\put(43,66){\makebox(0,0){\tiny{constraint}}}
\put(60,62){\line(-1,0){36}}
\qbezier(60,62)(70,62)(70,54)
\qbezier(70,54)(70,46)(87,46)
\put(87,46){\vector(1,0){0}}
\end{picture}}
\put(-3,4){\begin{picture}(0,0)(0,0)
\put(43,40){\makebox(0,0){\tiny{distributed}}}
\put(43,33){\makebox(0,0){\tiny{system}}}
\put(43,26){\makebox(0,0){\tiny{(alignment}}}
\put(43,19){\makebox(0,0){\tiny{diagram)}}}
\put(60,30){\line(-1,0){36}}
\qbezier(60,30)(66,30)(66,25)
\qbezier(66,25)(66,20)(78,20)
\put(78,20){\vector(1,0){0}}
\end{picture}}
\put(87,22){\begin{picture}(75,50)(0,0)
\setlength{\unitlength}{0.75pt}
\put(33,44){\makebox(0,0){\tiny{optimal}}}
\put(33,36){\makebox(0,0){\tiny{channel}}}
\put(33,6){\makebox(0,0){\tiny{channel}}}
\put(80,50){\makebox(0,0)[l]{\tiny{colimit}}}
\put(80,25){\makebox(0,0)[l]{\tiny{refinement}}}
\put(80,0){\makebox(0,0)[l]{\tiny{core}}}
\put(6,4)  {\vector(3,2) {65}}
\put(6,46) {\vector(3,-2){65}}
\put(0,45) {\vector(0,-1){40}}
\put(75,45){\vector(0,-1){40}}
\put(5,50) {\vector(1,0) {65}}
\put(5,0)  {\vector(1,0) {65}}
\put(0,0){\circle*{4}}
\put(0,50){\circle*{4}}
\put(75,0){\circle*{4}}
\put(75,50){\circle*{4}}
\qbezier[44](0, 5)(58,41)(75,50)
\qbezier[44](0,10)(58,42)(75,50)
\qbezier[44](0,15)(58,43)(75,50)
\qbezier[44](0,20)(58,44)(75,50)
\qbezier[44](0,25)(58,45)(75,50)
\qbezier[44](0,30)(58,46)(75,50)
\qbezier[44](0,35)(58,47)(75,50)
\qbezier[44](0,40)(58,48)(75,50)
\qbezier[44](0,45)(58,49)(75,50)
\qbezier[44](0,45)(58,9)(75,0)
\qbezier[44](0,40)(58,8)(75,0)
\qbezier[44](0,35)(58,7)(75,0)
\qbezier[44](0,30)(58,6)(75,0)
\qbezier[44](0,25)(58,5)(75,0)
\qbezier[44](0,20)(58,4)(75,0)
\qbezier[44](0,15)(58,3)(75,0)
\qbezier[44](0,10)(58,2)(75,0)
\qbezier[44](0, 5)(58,1)(75,0)
\qbezier(-15,25)(-15, 62)(0, 64)
\qbezier( 15,25)( 15, 62)(0, 64)
\qbezier(-15,25)(-15,-12)(0,-14)
\qbezier( 15,25)( 15,-12)(0,-14)
\end{picture}}
\end{picture}
\end{tabular}
\end{center}
\caption{Distributed System}
\label{distributed:system}
\end{figure}

The semantic integration of ontologies \cite{kent:dagstuhl}, \cite{goguen:06} 
can be represented by alignment and unification (Figure~\ref{distributed:system}): 
aligning a distributed system of ontologies 
by building a suitable diagram of logics and 
unifying the distributed system of ontologies along a channel 
covering the underlying diagram of structures.
The logics in the alignment diagram represent the individual ontologies, and 
the morphisms between logics in the alignment diagram represent the semantic alignment constraints. 
An example of semantic alignment constraints is 
the representation of an equivalent pair of types 
in two ontologies being aligned 
by a single type in a mediating ontology, 
with two mappings from this mediating type back to the equivalent pair of types. 
The alignment diagram represents a semantically constrained distributed system of ontologies, 
with individual logics representing parts of the system.
Any covering channel over the underlying diagram of structures
has a core that represents the whole system in some respect.
Unification forms a covering channel of logics that connects the distributed system to the fusion logic
--- the meet, 
in the logic fiber over the underlying core,
of the direct image of the diagram of logics 
along the underlying channel of structures.

\begin{principle}{Structure}
Information flow crucially involves structures of the world.
\emph{(This is the second principle of Information Flow \cite{barwise:seligman:97},
abstracted from classifications to structures. 
A classification is just one example of a structure.)}
\end{principle}

\noindent
This principle motivates 
the use of 
structures as the indexing objects for the (local) logics 
that incorporate the regularities of a distributed system
and 
the use of structure morphisms as the indexing links for the morphisms of (local) logics
that incorporate the information flow of regularities of a distributed system.

A distributed system 
$\mathcal{A} : \mathrmbf{I} \rightarrow \mathrmbf{Struc}$
consists of an indexed family 
$\{ \mathcal{A}_i \mid i \in |\mathrmbf{I}| \}$
of structures together with an indexed family 
$\{ \mathcal{A}_e : A_{i} \rightarrow A_{j} \mid (e : i \rightarrow j) \in \mathrmbf{I} \}$
of structure morphisms;
that is,
a distributed system is a diagram
in the structure category $\mathrmbf{Struc}$.
We think of the component structures $\mathcal{A}_i$ 
as being parts the the system.
We would also like to represent the whole system as a structure,
where we might have different representative structures for different purposes.
The theory of part-whole relations is called mereology.
It studies how parts are related to wholes, 
and how parts are related to other parts within a whole.
In a distributed system, 
the part to part relationships are modeled by the structure morphisms $\mathcal{A}_e : A_{i} \rightarrow A_{j}$.
In Information Flow,
we can model the whole as a structure $C$ 
and model the part-whole relationship between some part $A$ and the whole
with a structure morphism $g : A \rightarrow C$. 


\subsection{Information Channels}

An information channel $\mathcal{C} : A \stackrel{g}{\Rightarrow} \Delta(C)$
over a world (category of structures) $\mathrmbf{Struc}$
consists of an $I$-indexed family
$\mathcal{C} = {\{ g_i : A_i \rightarrow C \}}_{i \in I}$
of structure morphisms with a common target structure $C$,
called the core of the channel.
A channel $\mathcal{C} : A \stackrel{g}{\Rightarrow} \Delta(C)$
covers a distributed system
$\mathcal{A} : \mathrmbf{I} \rightarrow \mathrmbf{Struc}$
when $I = |\mathrmbf{I}|$
and the channel component morphisms commute with the distributed system morphisms
$A_{i} \stackrel{g_i}{\rightarrow} C
= A_{i} \stackrel{\mathcal{A}_e}{\rightarrow} A_{j} \stackrel{g_j}{\rightarrow} C$
for $e : i \rightarrow j$ in $\mathrmbf{I}$.
A covering channel
$\mathcal{C} : \mathcal{A} \stackrel{g}{\Rightarrow} \Delta(C)$
is essentially a cocone over diagram $\mathcal{A}$.
For any two covering channels
$\mathcal{C} : \mathcal{A} \stackrel{g}{\Rightarrow} \Delta(C)$
and
$\mathcal{D} : \mathcal{A} \stackrel{h}{\Rightarrow} \Delta(D)$
over the same distributed system $\mathcal{A}$,
a refinement (mediating morphism)
is a structure morphism between cores
$r : C \rightarrow D$
that commutes with the channel component morphisms
$A_{i} \stackrel{g_i}{\rightarrow} C \stackrel{r}{\rightarrow} D
= A_{i} \stackrel{h_j}{\rightarrow} C$
for $i \in |\mathrmbf{I}|$.
A channel $\mathcal{C}_{\mathrm{opt}} : \mathcal{A} \stackrel{\iota}{\Rightarrow} \Delta(\coprod\mathcal{A})$ 
is a minimal cover of a distributed system $\mathcal{A}$
when it covers $\mathcal{A}$ and for any other covering channel $\mathcal{D}$
there is a unique refinement $[\mathcal{D}] : \coprod\mathcal{A} \rightarrow D$
from $\mathcal{C}_{\mathrm{opt}}$ to $\mathcal{D}$. 
A minimal cover is essentially a colimiting cocone over diagram $\mathcal{A}$.
Any two minimal covers are isomorphic.

Information flow has two concerns with respect to channels:
(1) given a distributed system and some viewpoint (scientific, technological, social, etc.),
how should the whole system be modeled; and
(2) how does the natural logic of one component part of a system affect another component part.
The first concern, 
realizing a channel core, 
can have several solutions.
An optimal solution, the colimit, is discussed in the section on cocompleteness and cocontinuity. 
The second concern, 
involving distributed inference rules, local logics and information flow,
is discussed in other succeeding sections.

\begin{principle}{Connection}
It is by virtue of regularities among connections 
that information about some components of a distributed system carries information about other components.
\emph{(This is the third principle of Information Flow \cite{barwise:seligman:97}.)}
\end{principle}

\noindent
This principle motivates the use of logics over structures,
which lift theories over languages,
to represent information flow over covering channels of a distributed system.
For a simple example of information flow,
consider two component parts 
$A_{i} \stackrel{g_i}{\rightarrow} C \stackrel{g_j}{\leftarrow} A_{j}$
with underlying language morphisms
$\gamma_i$ and $\gamma_j$
that are connected to the core structure $C$
by being essentially projections
$A_i \cong \mathrmbfit{struc}(\gamma_i)(C)$ and $\mathrmbfit{struc}(\gamma_j)(C) \cong A_j$.
Then $A_i$'s satisfying sentence $a_i$ carries the information that $A_j$ satisfies sentence $a_j$, 
relative to the channel $\mathcal{C}$,
if 
the translation $\mathrmbfit{sen}(\gamma_i)(a_i)$ 
entails 
the translation $\mathrmbfit{sen}(\gamma_j)(a_j)$ 
in the theory $\mathrmbfit{max}(C)$.


\subsection{Inference Rules}


In the section we paraphrase the discussion in the first part of \cite{barwise:seligman:97}.
To see how unsound and incomplete logics arise in reasoning about distributed systems,
we consider the diagram
\begin{center}
$P \stackrel{p}{\longrightarrow} C \stackrel{d}{\longleftarrow} D$ 
\end{center}
called a binary channel.
This consists of 
a proximal structure   $P$ 
a distal structure     $D$ 
and 
a connecting structure $C$.
We think of this binary channel as representing a distributed system
having proximal part $P$, distal part $D$ and whole (or core) $C$.

We  are interested in discovering 
what kind of theory of the distal part is available to someone with complete knowledge of the proximal part.
The diagram suggests breaking the problem up into two parts,
the problem of going directly along $p$ from proximal component $P$ to core component $C$, and
the problem of going inversely along $d$ from core component $C$ to distal component $D$.
We can discuss both steps at once by considering a single structure morphism 
$f : M_1 \rightarrow M_2$
(in the above, $f$ can be 
either $p : P \rightarrow C$ 
    or $d : C \rightarrow D$).
Image someone who wants to reason about one side by using the induced theory of the other side.

Consider the following ``rules of inference''
along an structure morphism $f : M_1 \rightarrow M_2$
with underlying language morphism
$\mathrmbfit{lang}(f) = \sigma : \Sigma_1 \rightarrow \Sigma_2$.
The first says that from any source sentence $s_1 \in \mathrmbfit{sen}(\Sigma_1)$
we can infer the target sentence $\mathrmbfit{sen}(\sigma)(s_1) \in \mathrmbfit{sen}(\Sigma_2)$.
The second is read similarly.
\begin{center}
{\small \begin{tabular}{c@{\hspace{30pt}}c}
\begin{tabular}{l@{\hspace{5pt}}l}
$f$-{\bfseries Intro:}
& \begin{tabular}{c} $s_1$ \\ \hline $\mathrmbfit{sen}(\sigma)(s_1)$ \end{tabular}
\end{tabular}
&
\begin{tabular}{l@{\hspace{5pt}}l}
$f$-{\bfseries Elim:}
& \begin{tabular}{c} $\mathrmbfit{sen}(\sigma)(s_1)$ \\ \hline $s_1$ \end{tabular}
\end{tabular}
\end{tabular}}
\end{center}
The first rule allows us to move along the structure morphism $f$ 
from a source sentence to a target sentence,
whereas the second rule allows us to move in the opposite direction.
These inference rules have very important properties.
First consider the preservation of validity and nonvalidity.

A rule preserves validity 
when it leads from premise constraints to conclusion constraints.
The $f$-{\bfseries Intro} rule preserves validity:
if the premise $s_1$ is valid in $M_1$,
$M_1 \models_{\Sigma_1} s_1$,
then the conclusion $\mathrmbfit{sen}(\sigma)(s_1)$ is valid in $M_2$,
$M_2 \models_{\Sigma_2} \mathrmbfit{sen}(\sigma)(s_1)$.
This follows immediately from the definition of structure morphism.
The $f$-{\bfseries Elim} rule does not preserve validity.
It is possible to have a constraint $\mathrmbfit{sen}(\sigma)(s_1)$ of $M_2$,
$M_2 \models_{\Sigma_2} \mathrmbfit{sen}(\sigma)(s_1)$,
such that $s_1$ has counterexample $M_1$,
$M_1 \not\models_{\Sigma_1} s_1$.
However,
$M_1$ is not a counterexample of $s_1$
if $M_1$ is a specialization of $\mathrmbfit{struc}(\sigma)(M_2)$,
$M_1 \leq_{\Sigma_1} \mathrmbfit{struc}(\sigma)(M_2)$;
that is,
the $f$-{\bfseries Elim} rule is sound when $M_1 \cong_{\Sigma_1} \mathrmbfit{struc}(\sigma)(M_2)$.
In particular, the $f$-{\bfseries Elim} rule preserves validity along structure-isomorphic structure morphisms.

A rule preserves nonvalidity 
when it leads from premise nonconstraints to conclusion nonconstraints.
The $f$-{\bfseries Elim} rule preserves nonvalidity:
if the premise $\mathrmbfit{sen}(\sigma)(s_1)$ is not valid in $M_2$,
$M_2 \not\models_{\Sigma_2} \mathrmbfit{sen}(\sigma)(s_1)$,
then the conclusion $s_1$ is not valid in $M_1$,
$M_1 \not\models_{\Sigma_1} s_1$.
This also follows immediately from the definition of structure morphism.
The $f$-{\bfseries Intro} rule does not preserve nonvalidity.
It is possible that $s_1$ has counterexample $M_1$,
$M_1 \not\models_{\Sigma_1} s_1$,
where $\mathrmbfit{sen}(\sigma)(s_1)$ is a constraint of $M_2$,
$M_2 \models_{\Sigma_2} \mathrmbfit{sen}(\sigma)(s_1)$.
However,
$\mathrmbfit{sen}(\sigma)(s_1)$ is not a constraint of $M_2$
if $M_1$ is a specialization of $\mathrmbfit{struc}(\sigma)(M_2)$,
$M_1 \leq_{\Sigma_1} \mathrmbfit{struc}(\sigma)(M_2)$;
that is,
the rule preserves nonvalidity when $M_1 \cong_{\Sigma_1} \mathrmbfit{struc}(\sigma)(M_2)$.
In particular, the $f$-{\bfseries Intro} rule preserves nonvalidity along structure-isomorphic structure morphisms.

Summarizing the above,
the rule of $f$-{\bfseries Intro} preserves validity, but not nonvalidity;
whereas,
the rule of $f$-{\bfseries Elim} preserves nonvalidity, but not validity.
In terms of distributed systems,
when using the $f$-{\bfseries Intro} rule
any constraint that holds for a component
translates directly to a constraint about the whole system,
and
when using the $f$-{\bfseries Elim} rule
any constraint about the whole system
translates inversely to a constraint of those parts
that really are (up to isomorphism) a component of the system structure.
Returning to the binary channel
\begin{center}
$P \stackrel{p}{\longrightarrow} C \stackrel{d}{\longleftarrow} D$ 
\end{center}
depicted previously,
we wanted to know what happens when we use the sound and complete theory 
$\mathrmbfit{max}(P) = \{ s \in \mathrmbfit{sen}(\Sigma_P) \mid P \models_{\Sigma_P} s \}$
of the proximal structure to reason about the distal structure.
On the one hand,
we have seen that $p$-{\bfseries Intro} preserves validity, but not nonvalidity;
so that the theory we obtain at $C$ may be sound
(any constraint of $P$ maps to a constraint of $C$), 
but not necessarily complete 
(we may have no sentence or only nonvalid sentences of $P$ 
mapping to a particular constraint of $C$
--- there may be constraints of $C$ that are missed).
On the other hand,
following $p$-{\bfseries Intro} by $d$-{\bfseries Elim} 
means that we lose our guarantee that the resulting distal theory is either sound or complete.
A sentence about the distal structure obtained from a sentence about the proximal structure
in this way
is guaranteed to apply when 
the distal structure is connected (up to isomorphism) to the proximal structure in the channel.


\begin{figure}
\begin{center}
\begin{tabular}{c@{\hspace{80pt}}c}
\\
\setlength{\unitlength}{0.95pt}
\begin{picture}(40,100)(0,0)
\put(15,102){\makebox(0,0)[r]{\scriptsize{$\top$}}}
\put(26,102){\makebox(0,0)[l]{\scriptsize{$= \langle \Sigma,M,\emptyset \rangle = \top$}}}
\put(17,50){\makebox(0,0)[r]{\scriptsize{$\mathrmbfit{log}(M)$}}}
\put(26,50){\makebox(0,0)[l]{\scriptsize{$= \langle \Sigma,M,\mathrmbfit{max}(M) \rangle$}}}
\put(15,-2){\makebox(0,0)[r]{\scriptsize{$\bot$}}}
\put(26,-2){\makebox(0,0)[l]{\scriptsize{$= \langle \Sigma,M,{\wp}\mathrmbfit{sen}(\Sigma) \rangle$}}}
\put(20,72){\makebox(0,0){\tiny{$\emph{sound}$}}}
\put(20,28){\makebox(0,0){\tiny{$\emph{complete}$}}}
\put(-75,50){\makebox(0,0)[l]{\footnotesize{\shortstack{$\mathrmbfit{fbr}(M)^{\mathrm{op}}$\\$\cong$\\$\mathrmbfit{fbr}(\Sigma)^{\mathrm{op}}$}}}}
\put(-30,50){\makebox(0,0)[l]{\small{$\left\{ \rule[15pt]{0pt}{30pt} \right.$}}}
\put(20,100){\circle*{3}}
\put(20,50){\circle*{3}}
\put(20,0){\circle*{3}}
\qbezier(-10,50)(-10,90)(20,100)
\qbezier(-10,50)(-10,10)(20,0)
\qbezier(50,50)(50,90)(20,100)
\qbezier(50,50)(50,10)(20,0)
\put(20,50){\line(-2,3){22}}
\put(20,50){\line(2,3){22}}
\qbezier[30](-2,83)(20,90)(42,83)
\qbezier[30](-2,83)(20,70)(42,83)
\put(20,50){\line(-2,-3){22}}
\put(20,50){\line(2,-3){22}}
\qbezier[30](-2,17)(20,10)(42,17)
\qbezier[30](-2,17)(20,30)(42,17)
\end{picture}
&
\setlength{\unitlength}{1.0pt}
\begin{picture}(100,60)(0,-20)
\put(0,40){\makebox(0,0){\small{$\mathrmbf{Log}$}}}
\put(50,40){\makebox(0,0){\small{$\mathrmbf{Th}$}}}
\put(0,0){\makebox(0,0){\small{$\mathrmbf{Struc}$}}}
\put(50,0){\makebox(0,0){\small{$\mathrmbf{Lang}$}}}
\put(100,0){\makebox(0,0){\small{$\mathrmbf{Pre}$}}}
\put(120,30){\makebox(0,0)[l]{\footnotesize{$\left. \rule[5pt]{0pt}{10pt} \right\} \emph{fibrations}$}}}
\put(139,11){\makebox(0,0)[r]{\scriptsize{$\emph{heterogenization}$}}}
\put(158,11){\makebox(0,0)[l]{\scriptsize{$\emph{homogenization}$}}}
\put(141,23){\vector(0,-1){24}}
\put(155,-1){\vector(0,1){24}}
\put(120,-8){\makebox(0,0)[l]{\footnotesize{$\left. \rule[5pt]{0pt}{10pt} \right\} \emph{indexed categories}$}}}
\put(-8,20){\makebox(0,0)[r]{\footnotesize{$\mathrmbfit{log}$}}}
\put(8,25){\makebox(0,0)[l]{\footnotesize{$\mathrmbfit{struc}$}}}
\put(8,15){\makebox(0,0)[l]{\footnotesize{$= \mathrmbfit{pr}_1$}}}
\put(54,20){\makebox(0,0)[l]{\footnotesize{$\mathrmbfit{lang}$}}}
\put(25,48){\makebox(0,0){\footnotesize{$\mathrmbfit{pr}_2$}}}
\put(25,-8){\makebox(0,0){\footnotesize{$\mathrmbfit{lang}$}}}
\put(75,-8){\makebox(0,0){\footnotesize{$\mathrmbfit{dir}$}}}
\put(50,-22){\makebox(0,0){\footnotesize{$\mathrmbfit{th}$}}}
\put(-5,10){\vector(0,1){20}}
\put(5,30){\vector(0,-1){20}}
\put(50,30){\vector(0,-1){20}}
\put(15,40){\vector(1,0){20}}
\put(15,0){\vector(1,0){20}}
\put(65,0){\vector(1,0){20}}
\put(33,17){\line(1,0){12}}
\put(33,17){\line(0,-1){12}}
\put(50,-10){\oval(90,12)[b]}
\put(95,-7){\vector(0,1){0}}
\end{picture}
\end{tabular}
\end{center}
\caption{The Logic Context}
\label{logic}
\end{figure}
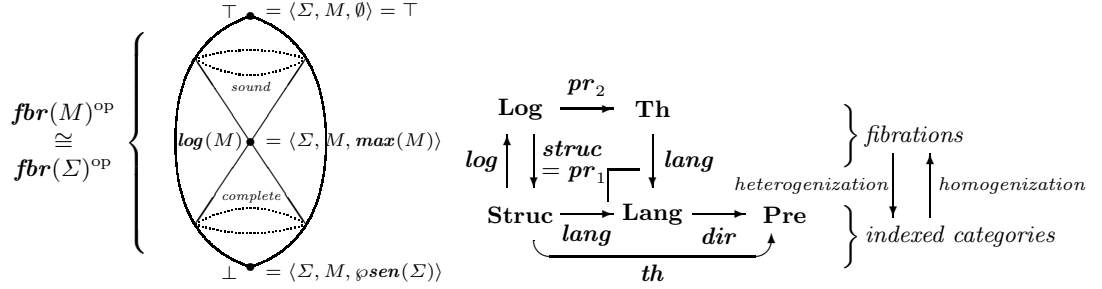

\subsection{Local Logics}

\begin{principle}{Channel}
The regularities of a given distributed system 
are relative to its analysis in terms of information channels.
\emph{(This is the fourth principle of Information Flow \cite{barwise:seligman:97}.)}
\end{principle}

\noindent
Paraphrasing and quoting \cite{barwise:seligman:97} (Lecture 12),
when ``reasoning about a distributed system with component'' parts of various kinds,
the component parts will typically be describe in quite different ways with different languages.
Along with these different languages 
``it is natural to think of each of the components as having its own logic'',
expressed in its own language.
``In this way, 
the distributed system'' of structures
``gives rise to a distributed system of local logics.
The interactions of the local logics reflect the behavior of the system as a whole.''
The concept of a (local) logic 
represents the regularities of a distributed system
and tracks what happens when we reason at a distance.

A (local) logic $L = \langle \Sigma, M, T \rangle$ consists of 
an indexing language $\Sigma$, 
a $\Sigma$-structure $M \in \mathrmbfit{struc}(\Sigma)$ and 
a $\Sigma$-theory $T \in \mathrmbfit{th}(\Sigma)$.
For any fixed structure $M$ with underlying language $\Sigma$,
the set of all logics with that structure is a preordered set under the theory order:
$\langle \Sigma, M, T_1 \rangle \leq \langle \Sigma, M, T_2 \rangle$ when $T_1 \leq T_2$.
This is the (opposite of the) fiber over $M$ with respect to the logic-to-structure projection functor
$\mathrmbfit{struc}$.
There are larger fibers.
For any fixed language $\Sigma$,
the set of all logics with that language is a preordered set under the structure and theory orders:
$\langle \Sigma, M_1, T_1 \rangle \leq \langle \Sigma, M_2, T_2 \rangle$ 
when $M_1 \leq_\Sigma M_2$ and $T_1 \leq_\Sigma T_2$.
This is the (opposite of the) fiber over $\Sigma$ with respect to the composite functor
$\mathrmbfit{struc} \circ \mathrmbfit{lang}$.
Any structure $M$ with underlying language $\Sigma$ 
induces the natural logic
$\mathrmbfit{log}(M) 
= \langle \Sigma, M, \mathrmbfit{max}(M) \rangle
= \langle \Sigma, M, M^\Sigma \rangle$. 
If two structures are ordered $M_1 \leq_\Sigma M_2$, 
then their logics are ordered $\mathrmbfit{log}(M_1) \leq_\Sigma \mathrmbfit{log}(M_2)$,
since $M_1 \leq_\Sigma M_2$
\underline{iff} $M_1^\Sigma \supseteq M_2^\Sigma$
\underline{iff} $M_1^\Sigma \leq_\Sigma M_2^\Sigma$.

A logic morphism $f : \langle \Sigma_1, M_1, T_1 \rangle \rightarrow \langle \Sigma_2, M_2, T_2 \rangle$
is a structure morphism $f : M_1 \rightarrow M_2$,
whose underlying language morphism $\sigma : \Sigma_1 \rightarrow \Sigma_2$
is also a theory morphism 
$\sigma : \langle \Sigma_1, T_1 \rangle \rightarrow \langle \Sigma_2, T_2 \rangle$. 
The category of logics $\mathrmbf{Log}$ 
has logics as objects and logic morphisms as morphisms.
It is describable (Figure~\ref{logic}) as 
either the pullback of the indexing functors for structures and theories,
or the Grothendieck construction of the dual indexed preorder
$\mathrmbfit{th} = \mathrmbfit{lang} \circ \mathrmbfit{dir} : \mathrmbf{Struc} \rightarrow \mathrmbf{Pre}$.
The projection functors
$\mathrmbfit{struc} = \mathrmbfit{pr}_1 : \mathrmbf{Log} \rightarrow \mathrmbf{Struc}$
and
$\mathrmbfit{pr}_2 : \mathrmbf{Log} \rightarrow \mathrmbf{Th}$
satisfy the pullback condition 
$\mathrmbfit{pr}_1 \circ \mathrmbfit{lang} = \mathrmbfit{pr}_2 \circ \mathrmbfit{lang}$.
The first pullback projection
is the fibration associated (homogenization) with the dual indexed preorder
$\mathrmbfit{th} : \mathrmbf{Struc} \rightarrow \mathrmbf{Pre}$.
This pullback projection
is the structure lift of the theory indexing functor
$\mathrmbfit{lang} : \mathrmbf{Th} \rightarrow \mathrmbf{Lang}$. 
The second pullback projection  
is the theory lift of the structure indexing functor
$\mathrmbfit{lang} : \mathrmbf{Struc} \rightarrow \mathrmbf{Lang}$.
Any structure morphism $f : M_1 \rightarrow M_2$,
with underlying language morphism $\sigma : \Sigma_1 \rightarrow \Sigma_2$,
induces the logic morphism $f : \mathrmbfit{log}(M_1) \rightarrow \mathrmbfit{log}(M_2)$
between natural logics.
This is well-defined,
since validity-preservation is equivalent to
$M_2^{\Sigma_2} \supseteq {\wp}\mathrmbfit{sen}(\sigma)(M_1^{\Sigma_1})$.
Hence,
there is a functor
$\mathrmbfit{log} : \mathrmbf{Struc} \rightarrow \mathrmbf{Log}$
(Figure~\ref{logic})
satisfying $\mathrmbfit{log} \circ \mathrmbfit{base} = 1_{\mathrmbf{Struc}}$.
A local logic $\langle \Sigma, M, T \rangle$ in the fiber over $M$
does not compare to the natural logic $\mathrmbfit{log}(M)$,
unless it is either sound or complete.

In general, logics may be neither sound nor (logically) complete.
A logic $\langle \Sigma, M, T \rangle$ is \emph{sound} 
when the structure $M$ satisfies the theory $T$ 
($T$ is valid in $M$),
$M \models_\Sigma T$;
equivalently, when $M^\Sigma \supseteq T$ or $M^\Sigma \supseteq T^\bullet$ or $M^\Sigma \leq_\Sigma T$.
A logic $\langle \Sigma, M, T \rangle$ is \emph{complete} 
when every sentence satisfied by $M$ is entailed by $T$;
that is, when $M \models_{\Sigma} t$ implies $T \vdash_{\Sigma} t$ for all sentences $t \in \mathrmbfit{sen}(\Sigma)$;
equivalently, when $T^\bullet \supseteq M^\Sigma$ or $T \leq_{\Sigma} M^\Sigma$;
equivalently, when $T \not\vdash_{\Sigma} t$ implies $M \not\models_{\Sigma} t$ 
for all sentences $t \in \mathrmbfit{sen}(\Sigma)$.
A logic $\langle \Sigma, M, T \rangle$ is sound (complete) 
\underline{iff} 
the identity (flat) structure morphism $1_M : M \rightarrow M$
with underlying identity language morphism $1_\Sigma : \Sigma \rightarrow \Sigma$
is a logic morphism 
$\eta_{\langle \Sigma, M, T \rangle} = 1_M : \langle \Sigma, M, T \rangle \rightarrow \mathrmbfit{log}(M)$ 
($\varepsilon_{\langle \Sigma, M, T \rangle} = 1_M : \mathrmbfit{log}(M) \rightarrow \langle \Sigma, M, T \rangle$).
The only sound and complete logics 
are those equivalent to the natural logic $\mathrmbfit{log}(M)$ for some structure $M$.
Sound logics form a reflective subcategory of all logics $\mathrmbf{Snd} \subseteq \mathrmbf{Log}$
with unit natural transformation $\eta : 1_{\mathrmbf{Snd}} \Rightarrow \mathrmbfit{struc} \circ \mathrmbfit{log}$.
Complete logics form a coreflective subcategory of all logics $\mathrmbf{Cmp} \subseteq \mathrmbf{Log}$
with counit natural transformation $\varepsilon : \mathrmbfit{log} \circ \mathrmbfit{struc} \Rightarrow 1_{\mathrmbf{Cmp}}$.

\subsection{Colimits}

\begin{principle}{Completeness/continuity}
The world is cocomplete and the description of the world is cocontinuous.
\end{principle}

\noindent
We assume the logical environment (Figure~\ref{logical:environment}) is cocomplete.
This means that it has 
a cocomplete
category of structures $\mathrmbf{Struc}$, 
a cocomplete category of logical languages $\mathrmbf{Lang}$, and
a cocontinuous fibration $\mathrmbfit{lang} : \mathrmbf{Struc} \rightarrow \mathrmbf{Lang}$.
Hence,
all colimits (universal constructions) of structures and languages are possible,
and the underlying language of the colimit of a diagram of structures 
is the colimit of the underlying diagram of languages.
All the examples of logical environments ($\mathtt{IFC}$, $\mathtt{EQ}$, $\mathtt{FOL}$) are cocomplete.

In 
approach advocated here, 
unpopulated ontologies (no world information and no semantics) are represented by theories
and the optimal semantic integration of unpopulated ontologies 
is represented by the colimit construction of theories,
whereas
populated ontologies (both world information and semantics) are represented by logics, 
and the optimal semantic integration of populated ontologies 
is represented by the colimit construction of logics. 
Colimits in the category of theories (logics) 
can be used to fuse together smaller theories (logics) to form larger ones. 
The colimit construction in the category of theories (logics) 
forms an optimal channel 
$\mathcal{C}_{\scriptscriptstyle\mathrm{opt}} 
: \mathcal{A} \stackrel{\iota}{\Rightarrow} \Delta({\coprod}\mathcal{A})$.
The fusion theory (logic) ${\coprod}\mathcal{A}$
is formed by information flow over the optimal channel: 
direct image flow followed by meet in the lattices of theories (logics). 
The colimit construction is based upon the colimit theorem, 
a powerful, general criterion for when such colimits of theories actually exist. 
It allows us to use for semantic integration, 
the same flow and lattice operators on logics-over-models as we do for theories-over-languages. 

\begin{theorem}
(Cocompleteness/Cocontinuity) \cite{goguen:burstall:92}
For any logical environment, 
the projection functors 
$\mathrmbfit{lang} : \mathrmbf{Struc}^{\flat} \rightarrow \mathrmbf{Lang}$ and
$\mathrmbfit{lang} : \mathrmbf{Th} \rightarrow \mathrmbf{Lang}$ reflect colimits. 
Hence, if the logical environment is cocomplete, 
then its category $\mathrmbf{Struc}^{\flat}$ of (flat) structures and structure morphisms 
and its category $\mathrmbf{Th}$ of theories and theory morphisms are cocomplete 
and the projection functors are cocontinuous. 
\end{theorem}


When the structure category $\mathrmbf{Struc}$ is cocomplete,
information flow in a channel 
$\mathcal{C} : \mathcal{A} \stackrel{g}{\Rightarrow} \Delta(C)$
covering a distributed system $\mathcal{A}$
can be 
factored through 
information flow in the optimal channel
followed by direct image along the unique mediating structure morphism
${\coprod}\mathcal{A} \stackrel{m}{\rightarrow} C$.

\subsection{Information Flow}

Information flow over a channel is defined as a two-step process: 
direct image flow along the component structure morphisms of the channel,
followed by the meet operation in the lattice of logics over the core. 

Given a structure morphism $f : M_1 \rightarrow M_2$
with underlying language morphism $\sigma : \Sigma_1 \rightarrow \Sigma_2$
and a source logic $L_1 = \langle \Sigma_1, M_1, T_1 \rangle$,
the \emph{direct image} is the target logic
$\mathrmbfit{dir}(f)(L_1) = \langle \Sigma_2, M_2, {\wp}\mathrmbfit{sen}(\sigma)(T_1) \rangle$.
The direct image is the greatest logic $L_2$ on the target structure $M_2$ 
such that $f$ is a logic morphism from $L_1$ to $L_2$;
that is,
$f : L_1 \rightarrow \mathrmbfit{dir}(f)(L_1)$ is a logic morphism; and
if $f : L_1 \rightarrow L_2$ is a logic morphism, then $L_2 \leq \mathrmbfit{dir}(f)(L_1)$.

\begin{proposition}
For any structure morphism $f : M_1 \rightarrow M_2$,
direct image preserves soundness:
if a source logic $L_1$ is sound, 
then the direct image logic $\mathrmbfit{dir}(f)(L_1)$ is also sound.
\end{proposition}

\paragraph{Proof:}
Assume the source logic $\langle \Sigma_1, M_1, T_1 \rangle$ is sound.
This means that 
$M_1^{\Sigma_1} \leq_{\Sigma_1} T_1$
(sentences inside the closure $T_1^\bullet$ are valid).
Since $f : M_1 \rightarrow M_2$ is an structure morphism,
with the underlying language morphism $\mathrmbfit{struc}(f) = \sigma : \Sigma_1 \rightarrow \Sigma_2$,
$\mathrmbfit{struc}(\sigma)(M_2)^{\Sigma_1} \leq_{\Sigma_1} M_1^{\Sigma_1}$.
Thus,
$\mathrmbfit{struc}(\sigma)(M_2)^{\Sigma_1} \leq_{\Sigma_1} T_1$.
By satisfaction invariance,
$\mathrmbfit{struc}(\sigma)(M_2)^{\Sigma_1} \leq_{\Sigma_1} T_1$
\underline{iff} 
$M_2^{\Sigma_2} \leq_{\Sigma_2} {\wp}\mathrmbfit{sen}(\sigma)(T_1)$,
this means that the direct image
$\mathrmbfit{dir}(f)(L_1) = \langle \Sigma_2, M_2, {\wp}\mathrmbfit{sen}(\sigma)(T_1) \rangle$
is sound.
\rule{5pt}{5pt}
\newline

\noindent
Given a structure morphism $f : M_1 \rightarrow M_2$
with underlying language morphism $\sigma : \Sigma_1 \rightarrow \Sigma_2$
and a target logic $L_2 = \langle \Sigma_2, M_2, T_2 \rangle$,
the \emph{inverse image} is the source logic
$\mathrmbfit{inv}(f)(L_2) = \langle \Sigma_1, M_1, \mathrmbfit{sen}(\sigma)^{-1}(T_2^\bullet) \rangle$,
where
$\mathrmbfit{sen}(\sigma)^{-1}(T_2^\bullet) 
= \{ t_1 \in \mathrmbfit{sen}(\Sigma_1) \mid T_2 \vdash_{\Sigma_2} \mathrmbfit{sen}(\sigma)(t_1) \}$.
The inverse image is the least logic $L_1$ on source structure $M_1$ 
such that $f$ is a logic morphism from $L_1$ to $L_2$;
that is,
$f : \mathrmbfit{inv}(f)(L_2) \rightarrow L_2$ is a logic morphism; and
if $f : L_1 \rightarrow L_2$ is a logic morphism, then $\mathrmbfit{inv}(f)(L_2) \leq L_1$.

\begin{proposition}
For any structure morphism $f : M_1 \rightarrow M_2$,
inverse image preserves completeness:
if a target logic $L_2$ is complete, 
then the inverse image logic $\mathrmbfit{inv}(f)(L_2)$ is also complete.
\end{proposition}

\paragraph{Proof.}
Assume target logic $\langle \Sigma_2, M_2, T_2 \rangle$ is complete.
This means that $T_2^\bullet \leq_{\Sigma_2} M_2^{\Sigma_2}$.
Hence,
$\mathrmbfit{sen}(\sigma)^{-1}(T_2^\bullet) \leq_{\Sigma_1} \mathrmbfit{sen}(\sigma)^{-1}(M_2^{\Sigma_2})$.
Since $f : M_1 \rightarrow M_2$ is an structure morphism
with language morphism $\mathrmbfit{struc}(f) = \sigma : \Sigma_1 \rightarrow \Sigma_2$,
we have $\mathrmbfit{sen}(\sigma)^{-1}(M_2^{\Sigma_2}) \leq_{\Sigma_1} M_1^{\Sigma_1}$.
Hence,
$\mathrmbfit{sen}(\sigma)^{-1}(T_2^\bullet) \leq_{\Sigma_1} M_1^{\Sigma_1}$.
This means that the inverse image
$\mathrmbfit{inv}(f)(L_2) = \langle \Sigma_1, M_1, \mathrmbfit{sen}(\sigma)^{-1}(T_2^\bullet) \rangle$
is complete.
\rule{5pt}{5pt}

\section{Conclusion}

We have defined logical environments,
semantic versions of institutions,
and have demonstrated how important concepts in {\ttfamily IF} theory,
such as distributed systems, channels and information flow, 
can be defined within logical environments.
Thus,
{\ttfamily IF} theory abstracts and extends {\ttfamily INS} theory.



\end{document}